\documentstyle[sprocl]{article}

\def\be{\begin{equation}}
\def\ee{\end{equation}}
\def\bea{\begin{eqnarray}}
\def\eea{\end{eqnarray}}

\begin{document}

\title{Dirac fields in a Bohm-Aharonov background and spectral boundary conditions}

\author{C. G. Beneventano, M. De Francia and E. M.
Santangelo}

\address{Departamento de F\'{\i}sica -
Facultad de Ciencias Exactas - \\ Universidad Nacional de La
Plata, Argentina}


\maketitle\abstracts{ We study the problem of a Dirac field in the
background of an Aharonov-Bohm flux string. We exclude the origin
by imposing spectral boundary conditions at a finite radius then
shrinked to zero. Thus, we obtain a behaviour of the
eigenfunctions which is compatible with the self-adjointness of
the radial Hamiltonian  and the invariance under integer
translations of the reduced flux. After confining the theory to a
finite region, we check the consistency with the index theorem,
and discuss the vacuum fermionic number and Casimir energy.}

\section{Setting of the problem}
\label{section-1}

We study the Dirac equation for a massless particle in four
dimensional Minkowski space, in the presence of a flux tube
located at the origin, i.e.,
\begin{equation}
\left( i \not\!\partial - \not\!\! A \right)\Psi = 0
\label{ec-1}\qquad \vec{H} = \vec{\nabla} \wedge \vec{A} =
\frac{\kappa}{r} \delta (r) \check{e}_z
\end{equation}
where $\kappa=\frac{\Phi}{2\pi}$ is the reduced flux.

As the gauge potential is $z-$independent, equation (\ref{ec-1})
can be decoupled into two uncoupled two-component
equations,\cite{Hagen} by choosing:
\begin{equation}
\gamma^0 = \left(
\begin{array}{cc}
  \sigma_3 & 0 \\
  0 & \sigma_3
\end{array}
\right) \gamma^1 = \left(
\begin{array}{cc}
  i \sigma_2 & 0 \\
  0 & i \sigma_2
\end{array}
\right)  \gamma^2 = \left(
\begin{array}{cc}
  -i\sigma_1 & 0 \\
  0 & i\sigma_1
\end{array}
\right) \gamma^3 = \left(
\begin{array}{cc}
  0 & i \sigma_1 \\
  i \sigma_1 & 0
\end{array}
\right)
\end{equation}

In order to avoid singularities, we will consider that only
$r>r_0$ is accessible, and take the limit $r_0\rightarrow 0$,which
is equivalent to having a punctured plane with the removed point
corresponding to the string position.

By taking $A_z = A_r =0, \, A_\theta =\frac{\kappa}{r}$, for $
r>r_0$, the Hamiltonian can be seen to be block-diagonal, with its
two-by-two blocks given by
\begin{equation}
H_\pm =\left(
\begin{array}{cc}
  0 & i e^{\mp i \theta} \left(\partial_r \pm {\rm B}\right)\\
   -i e^{\pm i \theta} \left(-\partial_r \pm {\rm B}\right) & 0
\end{array}
\right) \qquad {\rm B} = -\frac{i}{r} \partial_\theta
-\frac{\kappa}{r}
\end{equation}

It should be noticed that these two ``polarizations", which we
will label with $s=\pm1$, correspond to the two inequivalent
choices for the gamma matrices in 2+1 dimensions. From now on, we
will be working with $s=1$ (the case $s=-1$ can be studied in a
similar way, and explicit reference will be made to it whenever
necessary). In this case, we can write:
\begin{equation}
H_+ =\left(
\begin{array}{cc}
  0 & L^\dag \\
  L & 0
\end{array}
\right) \quad , {\rm with} \quad L=-i e^{i \theta}
\left(-\partial_r +{\rm B}\right) \quad L^\dag=i e^{-i \theta}
\left(\partial_r +{\rm B}\right)
\end{equation}
and its eigenfunctions:
\begin{equation}
\Psi_E = \left(\begin{array}{c}
  \varphi_E\left(r,\theta\right) \\
  \chi_E\left(r,\theta\right)
\end{array}
\right) \qquad,\quad{\rm satisfy:}\quad
\begin{array}{c}
  L \varphi_E = E \chi_E \\
  L^\dag \chi_E = E \varphi_E
\end{array}
\label{eq-9}
\end{equation}

Now, the two components in $\Psi_E$  have different $\theta$
dependence. In order to make this fact explicit, and to discuss
boundary conditions at $r=r_0$, we introduce \cite{Zhong}
\begin{equation}
\Psi_E = \frac{1}{\sqrt{r}} \left(
\begin{array}{c}
  e^{-i\frac{\theta}{2}} \varphi_{1E} \left(r,\theta\right)\\
  e^{i\frac{\theta}{2}} \chi_{1E} \left(r,\theta\right)
\end{array}
\right) \quad L_1 =-\partial_r + {\rm B} \quad L_1^\dag
=\partial_r + {\rm B}
\end{equation}
so that
\begin{equation}
L_1 \varphi_{1E} = i E \chi_{1E} \qquad L_1^\dag \chi_{1E} =-i E
\varphi_{1E} \label{eq-13}
\end{equation}

We expand $\varphi_{1E}$ and $\chi_{1E}$ in terms of
eigenfunctions of B, which are of the form:
\begin{equation}
e_n = e^{i\left(n+\frac{1}{2}\right)\theta} \qquad,\quad {\rm
with} \qquad
\lambda_n\left(r\right)=\frac{n+\frac{1}{2}-\kappa}{r} \quad,\quad
n\,\epsilon\, Z
\end{equation}
once the condition has been imposed that $\varphi_E$ and $\chi_E$
in equation (\ref{eq-9}) are single-valued in $\theta$.

Thus,we have
\begin{equation}
\varphi_{1E}\left(r,\theta\right) =\sum_{n=-\infty}^\infty
f_n\left(r\right) e^{i \left(n+\frac{1}{2}\right)\theta} \qquad
\chi_{1E}\left(r,\theta\right) =\sum_{n=-\infty}^\infty
g_n\left(r\right) e^{i\left(n+\frac{1}{2}\right)\theta}
\label{eq-16}
\end{equation}
which leads, for noninteger $\kappa$ ($\kappa=k+\alpha$, with $k$
the integer part of $\kappa$ and $\alpha$ its fractionary part) to

\begin{equation}
\Psi_E (r,\theta) = \sum_{n=-\infty}^\infty e^{i n\theta} \left(
\begin{array}{c}
  A_n\,J_{n-\kappa}\left(|E|r\right)+
  B_n\,J_{\kappa-n}\left(|E|r\right) \\
  -i\frac{|E|}{E}\left(A_n\,J_{n+1-\kappa}\left(|E|r\right)-
  B_n\,J_{\kappa-n-1}\left(|E|r\right)\right)
  e^{i \theta}
\end{array}
\right) \label{eq-18}
\end{equation}

(Of course, for integer $\kappa$, a linear combination of Bessel
and Neumann functions must be taken). Finally, for $s=-1$, the
upper and lower components of $\Psi_E$ interchange, and
$E\rightarrow-E$.

\section{Boundary conditions at the origin}
\label{section-2}

As is well known,\cite{Gerbert} the radial Dirac Hamiltonian in
the background of an Aharonov-Bohm gauge field requires a
self-adjoint extension for the critical subspace $n=k$. In fact,
imposing regularity of both components of the Dirac field at the
origin is too strong a requirement, except for integer flux.
Rather, one has to apply the theory of Von Neumann deficiency
indices, which leads to a one parameter family of allowed boundary
conditions, characterized by
\begin{equation}
i\,\lim_{r\rightarrow 0}\left(Mr\right)^{\nu +1} g_n(r)
\sin\left(\frac{\pi}{4}+\frac{\Theta}{2}\right)=
\lim_{r\rightarrow 0}\left(Mr\right)^{-\nu } f_n(r)
\cos\left(\frac{\pi}{4}+\frac{\Theta}{2}\right)
\end{equation}
with $\nu$ varying between $-1$ and $0$ ($\nu=-\alpha$ for $s=1$;
$\nu=\alpha-1$ for $s=-1$). Here, $\Theta$ parametrizes the
admissible self-adjoint extensions, and $M$, a mass parameter, is
introduced for dimensional reasons. Which of these boundary
conditions to impose depends on the physical situation under
study.

One possibility is to take a finite flux tube, ask for continuity
of both components of the Dirac field at finite radius and then
let this radius go to zero.\cite{Hagen} Thus, one of the possible
self-adjoint extensions is obtained, which corresponds to $\Theta=
\frac{\pi}{2} {\rm sgn}\left(\kappa\right)$. As pointed out, for
instance, in,\cite{sitenko96} this kind of procedure leads to a
boundary condition that breaks the invariance under
$\kappa\rightarrow\kappa+n$ ($n\,\epsilon\, Z$). Now, this is a
large gauge symmetry, which of course is singular when considering
the whole plane, but is not so when the origin is removed or,
equivalently, the plane has the topology of a cylinder.

To preserve the aforementioned symmetry we propose, instead, to
exclude the origin, by imposing spectral boundary conditions of
the Atiyah-Patodi-Singer(APS) type,\cite{aps} as defined
in,\cite{Zhong} at a finite radius $r_0$ and then letting
$r_0\rightarrow0$.

We consider the development in eq.(\ref{eq-16}) and, for $s=1$,
impose at $r=r_0$ :
\begin{equation}
f_n\left(r_0\right) = 0,\,{\rm for}\,
\lambda_n\left(r_0\right)\leq 0 \quad g_n\left(r_0\right) =
0,\,{\rm for}\,\lambda_n\left(r_0\right)>0 \label{eq-19}
\end{equation}

As is well known, imposing this kind of boundary condition is
equivalent to removing the boundary, by attaching a semi-infinite
tube at its position and then extending the Dirac equation by a
constant extension of the gauge field, while asking that zero
modes be square integrable \cite{Dowker6:1995} (except for
$\lambda_n=0$, where a constant zero mode remains, with a nonzero
lower component ).

After using the dominant behaviour of Bessel functions for small
arguments, and taking the zero radius limit, we have the following
result for the eigenfunctions in eq.(\ref{eq-18}):

\begin{eqnarray}
\Psi_E\left(r,\theta\right) & = &
 \sum_{n=-\infty}^k B_n\left(\begin{array}{l}
   J_{k+\alpha-n}\left(|E|r\right) e^{i n \theta} \\
   -i\frac{|E|}{E}J_{k+\alpha-n-1}\left(|E|r\right) e^{i\left(n+1\right) \theta}
 \end{array}\right)+ \nonumber \\ &&+
 \sum_{n=k+1}^\infty A_n\left(\begin{array}{l}
   J_{n-k-\alpha}\left(|E|r\right) e^{i n \theta} \\
   i\frac{|E|}{E}J_{n+1-k-\alpha}\left(|E|r\right) e^{i\left(n+1\right) \theta}
 \end{array}\right)\quad \alpha \geq \frac{1}{2}
 \label{eq-22}
 \end{eqnarray}

\begin{eqnarray}
\Psi_E\left(r,\theta\right) & = &
 \sum_{n=-\infty}^{k-1} B_n\left(\begin{array}{l}
   J_{k+\alpha-n}\left(|E|r\right) e^{i n \theta} \\
   -i\frac{|E|}{E}J_{k+\alpha-n-1}\left(|E|r\right) e^{i\left(n+1\right) \theta}
 \end{array}\right) \nonumber \\ &&+
 \sum_{n=k}^\infty A_n\left(\begin{array}{l}
   J_{n-k-\alpha}\left(|E|r\right) e^{i n \theta} \\
   i\frac{|E|}{E}J_{n+1-k-\alpha}\left(|E|r\right) e^{i\left(n+1\right) \theta}
 \end{array}\right)\quad\alpha<\frac{1}{2}
  \label{eq-24}
 \end{eqnarray}

 Notice that our procedure leads precisely to a self adjoint
 extension satisfying the condition of minimal irregularity (the
 radial functions diverge at $r\rightarrow 0$ at most as $r^{-p}$,
 with $p\leq \frac{1}{2}$). It corresponds to the values of the
 parameter $\Theta$ :
\begin{equation}
 \Theta =\left\{\begin{array}{rl}
   -\frac{\pi}{2} & \quad{\rm for}\, \alpha\geq\frac{1}{2} \\
   \frac{\pi}{2} & \quad{\rm for}\, \alpha<\frac{1}{2} \

 \end{array}\right.
 \end{equation}

 As shown in,\cite{manuel} $\Theta=\pm\frac{\pi}{2}$ are the
 only two possible values of the parameter which correspond to
 having a Dirac delta magnetic field at the origin. Moreover,
 this extension is compatible with periodicity in $\kappa$. In
 fact, the dependence on $k$ can be reduced to an overall phase
 factor in the eigenfunctions.

 As regards charge conjugation($\Psi_E\rightarrow \sigma_1
 {\Psi_E}^*;\,
 \kappa\rightarrow-\kappa$), it is respected by the eigenfunctions, except for
 $\alpha=\frac{1}{2}$. This is due to the already commented
 presence of a constant zero mode on the cylinder. However, for the
 representation $s=-1$ of $2\times 2$ Dirac matrices, APS boundary
 conditions must be reversed, for
 $\lambda\neq0$, as compared to (\ref{eq-19}), since the operator
 ${\rm B}$ changes into $-{\rm B}$. For $\lambda=0$ the lower
 component will be taken to be zero at $r_0$ which, as we will show
 later, allows for charge conjugation to be a symmetry of the whole
 model. In this case, the resulting extension corresponds to
 \begin{equation}
 \Theta =\left\{\begin{array}{rl}
   \frac{\pi}{2} & \quad{\rm for}\, \alpha\geq\frac{1}{2} \\
   -\frac{\pi}{2} & \quad{\rm for}\, \alpha<\frac{1}{2} \
 \end{array}\right.
 \end{equation}

 It is worth pointing that, for integer $\kappa=k$, our procedure
 leads (both for $s=\pm1$) to the requirement of regularity of both
 components at the origin.

\section{The theory in a bounded region} \label{section-3}

From now on, we will confine the Dirac fields inside a bounded
region, by introducing a boundary at $r=R$, and imposing there
boundary conditions of the APS type, complementary to the ones
considered at $r=r_0$. For $s=1$
\begin{equation}
f_n\left(R\right) = 0,\, {\rm for}\quad \lambda_n\left(R\right)>0
\qquad g_n\left(R\right) = 0,\,{\rm for}\quad
\lambda_n\left(R\right)\leq 0 \label{eq-25}
\end{equation}

We start by studying the zero modes of our theory, which are of
the form
\begin{equation}
\Psi_0\left(r,\theta\right)=\left(\begin{array}{l}
  e^{-i \frac{\theta}{2}}
  \sum_{n=-\infty}^\infty  A_n r^{n-k-\alpha}
  e^{i \left(n+\frac{1}{2}\right)\theta} \\
  e^{i \frac{\theta}{2}}  \sum_{n=-\infty}^\infty  B_n r^{k+\alpha-n-1}
  e^{i\left(n+\frac{1}{2}\right)\theta}
\end{array}
\right) \label{eq-26}
\end{equation}

It is easy to see that no zero mode remains after imposing the
boundary conditions at both the internal and external radius, even
without taking $r_0\rightarrow 0$. This is in agreement with the
APS index theorem.\cite{aps} In fact, according to such theorem
\begin{equation}
n_+ -n_- = {\cal A} + b\left(r_0\right)+b\left(R\right)
\end{equation}
where $n_+ (n_-)$ is the number of chirality positive (negative)
zero energy solutions, ${\cal A}$ is the anomaly, or bulk
contribution, and $b$ are the surface contributions coming from
both boundaries \cite{Zhong,spectral}
\begin{equation}
b\left(R\right) = \frac{1}{2}\left(h_R-\eta \left(R\right)\right)
\qquad b\left(r_0\right) =
\frac{1}{2}\left(\eta\left(r_0\right)-h_{r_0}\right)
\end{equation}
with $\eta(r)$ the spectral asymmetry of the boundary operator
${\rm B}$ and $h_r$ the dimension of its kernel.

In our case, the boundary contributions cancel. As regards the
volume part, it also vanishes for the gauge field configuration
under study, and we have $n_+ -n_- =0$, which is consistent with
the absence of zero modes. For $s=-1$ both boundary contributions
interchange, and identical conclusions hold regarding the index.

\bigskip

The nonzero energy spectrum can be determined by imposing (for
$s=1$) the boundary conditions (\ref{eq-25}) at $r=R$ on the
eigenfunctions in eqs.(\ref{eq-22}) and (\ref{eq-24}). Thus, one
gets:
\begin{equation}
E_{n,l}=\left\{
\begin{array}{l}
  \pm\frac{j_{n-\alpha,l}}{R},
  \, n \geq 1 \\
  \pm\frac{j_{n+\alpha,l}}{R},
  \, n\geq -1
\end{array}
\right.,\, \,\alpha\geq \frac{1}{2} \quad E_{n,l}= \left\{
\begin{array}{l}
  \pm\frac{j_{n-\alpha,l}}{R},
  \, n\geq 0\\
  \pm\frac{j_{n+\alpha,l}}{R},
  \, n\geq 0
\end{array}
\right. ,\,\,\alpha< \frac{1}{2} \label{eq-30}
\end{equation}
where $j_{\nu,l}$ is the $l$-th positive root of $J_\nu$. The same
spectrum results for $s=-1$.

For both $s$ values, the energy spectrum is symmetric with respect
to zero. This fact, together with the absence of zero modes
results in a null vacuum expectation value for the fermionic
charge \cite{Niemi1984}
\begin{equation}
\langle N \rangle_{+} =-\frac{1}{2}\left(n_+-n_-\right)= 0
\end{equation}

For the same reasons $\langle N \rangle_{-} =0$, so the total
fermionic number of the theory is null.

It is interesting to note that the contribution to the fermionic
number coming from $r_0$ coincides, for each $s$ value, with the
result presented for the whole punctured plane in
\cite{sitenko96}(for details, see\cite{hep}), except for
$\alpha=\frac{1}{2}$. This last fact is associated with charge
conjugation non invariance in each subspace. However, the sum of
both contributions cancels for all $\alpha$.
\bigskip

We go now to the evaluation of the Casimir energy which, in the
framework of the $\zeta$ regularization, is given by
\begin{equation}
E_C =-2\mu\left. \left(\mu R\right)^z \sum_\nu
\zeta_\nu\left(z\right)\right\rfloor_{z=-1} \label{eq-35}
\end{equation}
where the parameter $\mu$ was introduced for dimensional reasons.
Here, $\zeta_\nu\left(z\right)$ is the so-called partial zeta
function, defined as in reference.\cite{Leseduarte1996comm}

For the problem at hand,
\begin{equation}
\sum_\nu \zeta_\nu = \left\{
\begin{array}{ll}
  \sum_{n=-\infty}^\infty \zeta_{|n-\alpha|} \,+ \zeta_{\alpha-1} &
  \qquad {\rm for}\quad \alpha \geq\frac{1}{2} \\
  \sum_{n=-\infty}^\infty \zeta_{|n-\alpha|} \,+ \zeta_{-\alpha} &
  \qquad {\rm for}\quad \alpha <\frac{1}{2}
\end{array}
\right. \label{eq-36}
\end{equation}

Notice that, as a consequence of the invariance properties of the
imposed boundary conditions, the Casimir energy is both periodic
in $\kappa$ and invariant under $\alpha\rightarrow 1-\alpha$, as
well as continuous at integer values of $\kappa$.

For any value of $\kappa$, it is the sum of the energy
corresponding to a scalar field in the presence of a flux string
and subject to Dirichlet boundary conditions, plus a partial zeta
coming, for fractionary $\kappa$, from the presence of an
eigenfunction which is singular at the origin or, for integer
$\kappa$, from the duplication of $J_0$. Both contributions can be
studied following the methods employed
in.\cite{Leseduarte1996comm} So, here we won't go into the details
of such calculation.

The scalar field contribution presents a pole at $z=-1$, with an
$\alpha$-independent residue. However, the pole in the partial
zeta appearing in (\ref{eq-35}),(\ref{eq-36}) has an
$\alpha$-dependent residue. Due to this fact, and the consequent
need to introduce $\alpha$-dependent counterterms, an absolute
meaning cannot be assigned to the finite part of the Casimir
energy.

As a final remark, it is worth stressing that this approach to the
problem of self-adjointness at the origin is, to our knowledge,
the first proposal of a physical application of APS boundary
conditions in this context.

Contrary to the treatment of the origin, APS boundary conditions
were imposed at the exterior boundary for merely formal reasons
(the existence of an index theorem for this case). The  vacuum
energy under local (bag-like) external boundary conditions is at
present under study.

\section*{Acknowledgments}

The authors thank H. Falomir for many useful discussions and
comments. E.M.S. also thanks the organizers of this event for
their kind hospitality. This work was partially supported by
ANPCyT, under grant PMT-PICT 0421 and U.N.L.P., Argentina.

\section*{Bibliography}


\end{document}